\documentclass[10pt,twocolumn]{article}

\usepackage[letterpaper,margin=1in]{geometry}
\usepackage{amsmath,amssymb,amsfonts}
\usepackage{algorithmic}
\usepackage{graphicx}
\usepackage{textcomp}
\usepackage{hyperref}
\usepackage{xcolor}
\usepackage{listings}
\usepackage{float}
\usepackage{enumitem}
\usepackage{subfig}

\title{
\textbf{An AI Implementation Science Study to Improve Trustworthy Data in a Large Healthcare System} \\
\large\textit{Preprint version. This manuscript has been accepted at IEEE BHI 2025. 
This is the author-prepared version and not the final published IEEE version. 
The final version will appear in IEEE Xplore.}
}

\author{
\begin{tabular}{c}
Benoit L. Marteau$^{1}$, Andrew Hornback$^{1}$, Shaun Q. Tan$^{1}$, \\
Christian Lowson$^{2}$, Jason Woloff$^{2}$, May D. Wang$^{1}$ \\
\small
$^{1}$Georgia Institute of Technology, Atlanta, GA, USA \\
\small
$^{2}$Shriners Hospitals for Children, Tampa, FL, USA \\
\small Email: benoitmarteau@gatech.edu, ahornback6@gatech.edu, stan99@gatech.edu, \\
\small christian.lowson@shrinenet.org, jason.woloff@shrinenet.org, maywang@gatech.edu
\end{tabular}
}

\date{}

\begin{document}
\maketitle

\begin{abstract}
The rapid growth of Artificial Intelligence (AI) in healthcare has sparked interest in Trustworthy AI and AI Implementation Science, both of which are essential for accelerating clinical adoption. Yet, barriers such as strict regulations, gaps between research and clinical settings, and challenges in evaluating AI systems hinder real-world implementation. This study presents an AI implementation case study within Shriners Children's (SC), a large multisite pediatric system, showcasing the modernization of SC's Research Data Warehouse (RDW) to OMOP CDM v5.4 within a secure Microsoft Fabric environment. We introduce a Python-based data quality assessment tool compatible with SC's infrastructure, an extension of OHDSI's R/Java-based Data Quality Dashboard (DQD) that integrates Trustworthy AI principles using the METRIC framework. This extension enhances data quality evaluation by addressing informative missingness, redundancy, timeliness, and distributional consistency. We also compare systematic and case-specific AI implementation strategies for Craniofacial Microsomia (CFM) using the FHIR standard. Our contributions include a real-world evaluation of AI implementations, integration of Trustworthy AI in data quality assessment, and evidence-based insights into hybrid implementation strategies, highlighting the need to blend systematic infrastructure with use-case-driven approaches to advance AI in healthcare.
\end{abstract}


\section{Introduction}
\label{Intro}

Artificial Intelligence (AI) has made significant progress over the past decade, driven by advancements in technology and adoption. In healthcare, this has led to a growing emphasis on Trustworthy AI (TAI), aimed at mitigating uncertainty and improving data and model transparency, as well as AI Implementation Science, which identifies barriers and practical solutions to AI adoption in clinical settings.

However, implementing AI in healthcare remains challenging due to strict data privacy regulations and the multimodal nature of patient data, including time series, monitoring, imaging, genomics, and structured or unstructured Electronic Health Records (EHRs) \cite{acosta2022multimodal, bohr2020current, gerke2020ethical}. These challenges are amplified in large multisite healthcare systems. Current research addresses these issues by improving model generalizability through data standardization, federated learning, or foundation models \cite{alsaad2024multimodal, kumar2021data, chen2021personalized, antunes2022federated, he2024foundation}.

Given the choice between working on AI model implementation or data quality assessment and improvement, we prioritized building the foundations for high-quality data, recognizing that model performance depends on the quality of the input data. To support our research, we adopted two standards, the Fast Healthcare Interoperability Resources (FHIR) standard for data harmonization and the Observational Medical Outcomes Partnership (OMOP) Common Data Model (CDM) for data standardization \cite{FHIR}. 
We identify several critical potential weaknesses in current research that hinder the implementation and widespread adoption of AI in healthcare. There is a gap in AI research conducted in controlled environments with curated datasets, and real-world implementation \cite{el2025bridging}. Most existing frameworks prioritize model evaluation over implementation or data quality improvement, overlooking the nuanced challenges and opportunities involved in deploying AI in clinical practice \cite{you2025clinical}. The challenges lie in the multidimensionality of data quality evaluation, encompassing the adherence to technical standards, the fidelity of data in representing real-world phenomena, and the usefulness of data for both AI models and users. 
While tools like Observational Health Data Sciences and Informatics (OHDSI)’s Data Quality Dashboard (DQD) support structured evaluations within OMOP CDM, broader frameworks such as Measurement Process, Timeliness, Representativeness, Informativeness, Consistency (METRIC) offer more comprehensive but abstract guidance, leaving implementation details to users \cite{kahn2016harmonized, schwabe2024metric}. Although current AI Implementation Science emphasizes systematic, generalizable frameworks, these often fall short when applied to specific healthcare use cases, which require tailored approaches.

To address these gaps, we collaborated with Shriners Children’s (SC), a large multisite pediatric healthcare system with over 22 hospitals across North America. SC provides an ideal case study due to its diverse, multimodal data and its multiple specialties, including craniofacial disorders, burns, and orthopedics, offering unique case studies. Our key contributions are as follows:

\begin{itemize}
    \item We provide Real-World Evidence (RWE) of data infrastructure standardization and modernization from an AI Implementation Science study in a real-world, multisite, and multimodal healthcare system.
    \item We extend the OHDSI DQD standard data quality evaluation to include TAI approaches and the METRIC framework.
    \item We provide evidence-based details and insights into the differences and similarities between systematic and case study-specific implementations.
\end{itemize}

\section{Background}
\label{Back}
\subsection{SC Data Infrastructure}
SC established the Shriners Health Outcomes Network (SHONet) to build a Research Data Warehouse (RDW) following the OMOP CDM. The SHONet initiative began as a means to leverage SC's data, thereby enhancing SC's clinical efficacy studies and enabling its clinicians to conduct comprehensive patient cohort analyses. SC utilizes standardized Extract-Transfer-Load (ETL) processes to map its data from its Cerner Millennium and newer Epic System EHR systems. SC RDW is currently housed in a secure Microsoft (MS) Azure environment with more than $240$ Gigabytes (GB) of data and billions of data points, accessible only via Azure Virtual Machines (VMs) controlled via a Role-Based Access Control (RBAC) mechanism. 

Recently, SC adopted MS Fabric, enabling researchers to access real-world data within a secure environment that complies with the Health Insurance Portability and Accountability Act (HIPAA) \cite{FABRIC}. MS Fabric integrates various data services, such as Lakehouse data storage that leverages Spark DataFrames for large-scale data processing and analytics. Moreover, the MS Fabric notebook provides an interactive coding environment to enable data engineers and scientists to develop programs and AI models with direct access to the data. For this study, SC copied its RDW into an MS Fabric workspace environment, which we used to perform the various experiments and analyses. 

\begin{figure}[t]
    \centering    
    \includegraphics[width=0.95\linewidth]{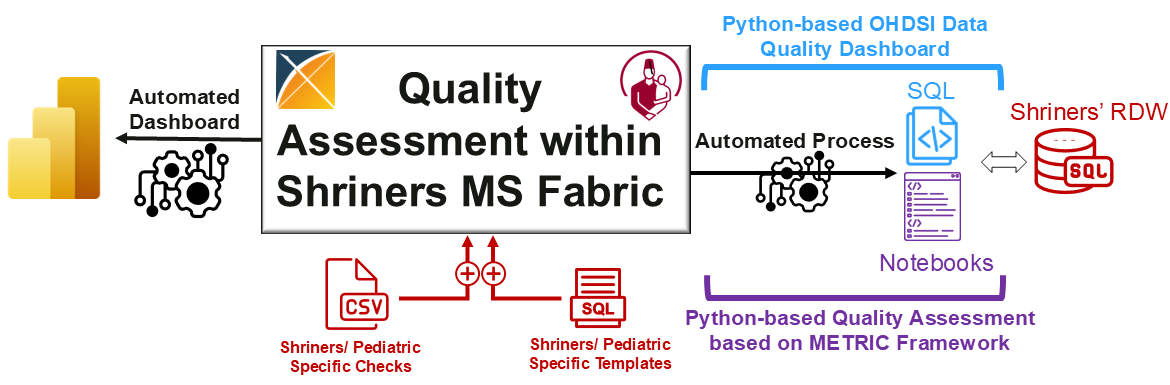}
    \caption{Overview of our adaptation and implementation of OHDSI Data Quality Dashboard (DQD) within Shriners Children's MS Fabric environment.}
    \label{fig:dqd}
\end{figure}

\subsection{OHDSI OMOP CDM}
OHDSI developed the OMOP CDM data standard to enable large-scale collaboration between healthcare institutions \cite{hripcsak2015observational}. This standard revolves around the OMOP CDM Concept Code ID, which represents various concepts encountered in medical practice, including procedures, measurements, drugs, devices, observations, and conditions. To support the OMOP CDM, OHDSI developed a standard vocabulary to harmonize the different classification and code systems that medical professionals use in their practice (e.g., International Classification of Diseases (ICD)-9, ICD-10, Systematized Nomenclature of Medicine - Clinical Terminology (SNOMED-CT), Current Procedural Terminology, 4th Edition (CPT4), etc) \cite{Athena}. OHDSI also developed a tool suite to facilitate the ETL process between source EHR data and OMOP CDM databases, or to evaluate the quality of an OMOP CDM database with the DQD \cite{kahn2016harmonized}. 

\subsection{HL7 FHIR}
FHIR is a modern interoperability standard developed by Health Level Seven (HL7) that enables the structured exchange of healthcare data using internet-based technologies. The standard defines a set of modular data components, called Resources, that represent common clinical concepts. These Resources, such as Patient, Observation, Medication, and Condition, can be accessed and manipulated through Representational State Transfer (RESTful) Application Programming Interfaces (API), typically in JavaScript Object Notation (JSON) or eXtensible Markup Language (XML) formats, allowing for flexible and scalable data integration across healthcare systems \cite{FHIR}.

\subsection{METRIC Framework}

The METRIC framework published by Schwabe et al. in 2024 plays a crucial role in the development of TAI systems by providing a structured framework to assess the quality of training data, with the assumption that high-quality and reliable data is key to robust and ethical AI systems, infering that we can't have TAI without trustworthy data \cite{schwabe2024metric}. The authors developed this framework based on a systematic review of scientific studies to increase the adoption of AI in healthcare, providing a set of guidelines for AI scientists, engineers, end-users, and regulators. However, this framework does not provide any specific software or tools contrary to OHDSI DQD, and therefore, its implementation relies on the developer. 
This framework comprises five dimensions: 1) \textbf{\textit{MEasurement Process}} measures the uncertainty related to the data acquisition, from sensors to human-induced error and source credibility; 2) \textbf{\textit{Timeliness}} ensures that the data is up to date with the latest knowledge and standard (e.g., ICD9 vs. ICD10); 3) \textbf{\textit{Representativeness}} measures how well the data represents a target population; 4) \textbf{\textit{Informativeness}} evaluate the amount of information represented by the data (e.g., data redundancy, data missingness); and 5) \textbf{\textit{Consistency}} measures the consistency of a dataset concerning standards and free of contradictions.

\subsection{Case Study: Craniofacial Microsomia} 
Craniofacial Microsomia (CFM) is a complex congenital condition characterized by the underdevelopment of the ear, mandible, and associated facial structures. Care and treatment often require long-term, multidisciplinary care across psychosocial, surgical, and other domains. Given the variability in phenotypic presentation and treatment trajectories, managing CFM presents significant challenges for care coordination, data standardization, and clinical decision-making. One crucial task is evaluating the impact of surgeries and CDM on patient mental health. This makes CFM an ideal use case for exploring how AI implementation science, driven by FHIR, can support personalized care planning, automate data integration across specialties, and potentially enhance longitudinal tracking of outcomes. 
\section{AI Implementation Science}
\label{Methods}

AI implementation science is an emerging field that focuses on bridging the gap between the development of AI models and their practical and ethical integration into real-world healthcare settings. Unlike traditional AI research, which often emphasizes model performance in controlled environments, it focuses on how these tools are adopted, utilized, and evaluated in complex clinical workflows. Another specificity of AI implementation science is its heavy reliance on engineering. However, unlike AI engineering, which focuses on building the tools, AI implementation science is the systematic study of how to integrate evidence-based practices/ interventions/ approaches into real-world practice to bridge the gap of the well-documented “know-do-gap” \cite{wray2023bridging}.

\subsection{Modernization and Evaluation of SC RDW}
SC developed its RDW around 2015, coinciding with the development of the OMOP CDM versions 5.1 and 5.2. Due to the complexity associated with developing new ETL pipelines, SC RDW still adheres to the same version, thereby limiting its potential use for collaborative research or the implementation of peer-reviewed tools \cite{koscielniak2022shonet, reinecke2021usage, marteau2024improving}. Moreover, SC personalized some tables to fit their needs, specifically with data related to the ETL process version, care site specialty, a table specific to Patient-Reported Outcome Measurements (PROM) Observations, and pain mitigation medications.

To modernize SC RDW to the latest OMOP CDM version 5.4, we started by mapping SC's RDW tables and columns to version 5.4. We then identified the relevant functions and logic of the DQD developed by OHDSI, which includes the generation of SQL scripts based on templates. We then selected Python as the programming language as it is natively implemented and supported by multiple environments (such as MS Fabric or Databricks). We then implemented our version of the DQD within MS Fabric, validating the SQL scripts generated by our implementation with the SQL scripts generated by the original DQD by OHDSI. We ultimately evaluated the quality of SC RDW before and after modernization. 
Essentially, we tried to answer the following Research Questions (RQs):
\begin{itemize}
    \item \textbf{\textit{RQ1}}: How does modernizing SC OMOP CDM databases influence its data quality assessment using OHDSI DQD?
\end{itemize}

\subsubsection{AI Implementation Challenges and Opportunities}
OHDSI developed its tools using the R and Java programming languages, both of which are technically supported by MS Fabric. However, these tools require a specific version of Java that is incompatible with MS Fabric, presenting an implementation barrier. Previous research either had the option to create a whole infrastructure compatible with this version of Java or manually run the tools on the database \cite{lima2019transforming, marteau2024improving}. These tools usually comprise three parts: a part that uses templates to generate SQL scripts, a part that interacts with the database, and a part that acts as a dashboard with the use of web applications. We propose an alternative solution to the current OHDSI R-based implementation: converting the first part (SQL generation) to Python to be used as a package, while leaving the second and third parts as use-case dependent implementations. Specifically, we allow the user to change the interaction mechanism and dashboard visualization easily. This choice is based on the dependence between the interaction mechanism and dashboard visualization on the environment. For example, MS Fabric has specific APIs to interact with the databases, and we used Power BI to create our dashboard. We acknowledge that this still leaves some implementation to the user; however, this should be mitigated over time as more researchers implement different interaction mechanisms and dashboard visualizations across different environments and share them.  We provide an overview of our implementation and conversion of OHDSI's DQD in  \textbf{Fig. \ref{fig:dqd}}. 
We validated our implementation by comparing the SQL scripts generated and results obtained with the original OHDSI DQD. This iterative process enabled us to identify and fix code and logic mistakes in our implementation. Our current implementation yields the same SQL scripts and results as the R-based DQD.
We then integrated our converted code into an automated pipeline within SC MS Fabric and created an interactive dashboard that regularly monitors SC RDW quality over time. 

\subsection{TAI-based Implementation Evaluation}
We extended OHDSI's DQD with approaches inspired by TAI research, such as the METRIC framework. We believe that such a framework proposes a natural extension of OHDSI DQD, with additional evaluation dimensions. The primary challenge lies in applying these abstract evaluation concepts to real-world data. This is why in this study, we selected and implemented four quality assessments based on evaluation dimensions from the METRIC framework. Specifically, we selected Informative Missingness, Timeliness, and Distribution Consistency, as we could translate these evaluation concepts into specific research questions:

\begin{itemize}
    \item \textbf{\textit{RQ2 (Informative Missingness)}}: Are missing data presenting a specific pattern based on their type (e.g., procedure, condition) or hospital site?
    \item \textbf{\textit{RQ3 (Timeliness)}}: Are all unique source data mapped to the same concepts (e.g., are mappings of the same concept different based on the version of the code (ICD9 vs. ICD10))?
    \item \textbf{\textit{RQ4 (Distribution Consistency)}}: Is data distribution uniform across the different hospital sites?
\end{itemize}

We think that not all the assessments proposed by the METRIC framework can be applied to a systematic implementation evaluation. For example, evaluating Informativeness and Representativeness partially requires expertise and is case study dependent (e.g., what is the target population). The measurement process also requires manual expert investigation, notably to understand the potential causality chain that led to a device or human-induced error. 
That being said, some of these limitations to systematic evaluations can be overcome by using multiple data sources as a reference. For example, the application of Natural Language Processing (NLP) techniques to clinical notes may help mitigate human-induced errors and enhance credibility by providing multiple sources, as demonstrated by previous studies \cite{wang2021covid}. 

\begin{figure}[htbp!]
    \centering    
    \includegraphics[width=0.92\linewidth]{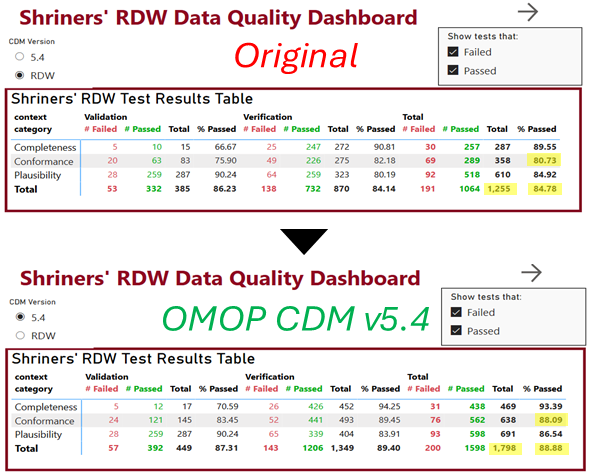}
    \caption{We modernized Shriners Children's Research Data Warehouse OMOP CDM database, and observed an improvement of the OHDSI DQD assessment, with more tests performed post-modernization.}
    \label{fig:rq2}
\end{figure}

\subsection{Systematic vs. Case Study}

To evaluate the benefits of our systematic approaches in a specific case study, we chose to work on a case study on CFM.
For this case study, we had access to only a subset of SC's RDW, which limited our ability to perform all the analyses that could be conducted using systematic approaches, notably regarding multisite analysis. However, this use case enabled us to assess the usefulness of the data, using AI models to analyze the impact of the patients' diagnoses and procedures on their mental health. This means that for our case study, we first had to define a cohort definition to retrieve the a priori relevant data with the help of clinicians. Moreover, clinicians were heavily involved in the data quality assessment and pre-processing, notably in our case study, providing important insights on how codes were generated. Specifically, the fact that multiple procedure codes are generated for a single surgery, the change of code vocabulary between ICD-9 and ICD-10 in 2015, and the evaluation of the relevance of the procedure and psychiatric diagnosis to the case study.

We also leveraged this case-study to evaluate the impact of data harmonization on AI performance. The AI model was trained using patients' diagnoses to classify whether they had a psychiatric-related diagnosis or not. We collaborated with clinicians to generate the list of relevant psychiatric diagnoses. We then identified the diagnoses and procedure codes of every patient, and created one-hot encoding as our input features. This means that we have as many features as different diagnoses and procedure codes, using "1" representing a code linked with the patient. We repeated the operation using the source codes (e.g., ICD, SNOMED, CPT4, ...) and then using the harmonized OMOP CDM concept codes. We present the results of the AI performance using either source or OMOP CDM concept codes in the next section.

This case-study enabled us to answer the following questions:
\begin{itemize}
    \item \textbf{\textit{RQ5}}: Does Data Harmonization impact AI model performance? Due to the use of multiple similar vocabulary (e.g., ICD9 and ICD10), using the OMOP CDM concept codes to harmonize the data reduices the effective number of unique source codes required to represent the data (which means that two source codes in different vocabulary represent the same concept, and therefore are mapped to the same OMOP CDM concept code).
    \item \textbf{\textit{RQ6}}: Does reducing the number of concept code to represent a data improves AI model performance (using OMOP CDM concept relationship by grouping OMOP CDM concept codes together in supersets)?
\end{itemize}

Ultimately, we focused on FHIR for the case study, as we believe this standard to be more appropriate for the concrete adoption of AI, not only with EHR data, but with multimodal data. Indeed, although the OMOP CDM is more suitable for storing observational EHR data, FHIR enables the creation of interactive and user-friendly web applications that can directly interact with the database, rather than merely visualizing it (e.g., through dashboards). To create the FHIR resources, we determined the exact nature of the data and how it could be converted into FHIR for AI-based integration with FHIR applications. The study dataset integrated patient demographics, clinical conditions, and surgical procedures, all mapped to HL7® FHIR® resources. The Patient resource captured demographics, using standard extensions for race and ethnicity. The Condition resource encoded diagnoses using ICD-10 and OMOP terminologies. The Procedure resource included both standardized OMOP codes for analytics and local source codes as custom URIs to ensure data fidelity. All Condition and Procedure records were linked to the corresponding Patient resource via the subject field.

\section{Results}
\label{Results}

\subsection{Improving SC Data Infrastructure}

Our implementation of the OMOP CDM v5.4 was partially successful, as we were able to map SC's existing RDW to OMOP CDM v5.4. However, numerous data points are missing, notably because the OMOP CDM v5.4 is more comprehensive than previous versions, resulting in some fields being left blank.  
We observed that the modernization had a positive impact according to the DQD, with a general quality test success rate improvement of 4\% (from 84.78\% to 88.88\%), and 8\% conformance improvement (from 80.73\% to 88.09\%), validating our \textbf{RQ1}. However, we were not able to achieve 100\% conformance, notably because multiple data points were in the wrong table/ category (e.g., an observation in the procedure table), meaning that further investigation is required. We also developed our dashboard using Power BI, showing pre- and post-modernization DQD results in a user-friendly and accessible interactive application, which we show \textbf{Fig. \ref{fig:rq2}}.

\subsection{Implementation Evaluation Analysis}

\begin{figure}[htbp!]
    \centering    
    \includegraphics[width=0.95\linewidth, height=4cm]{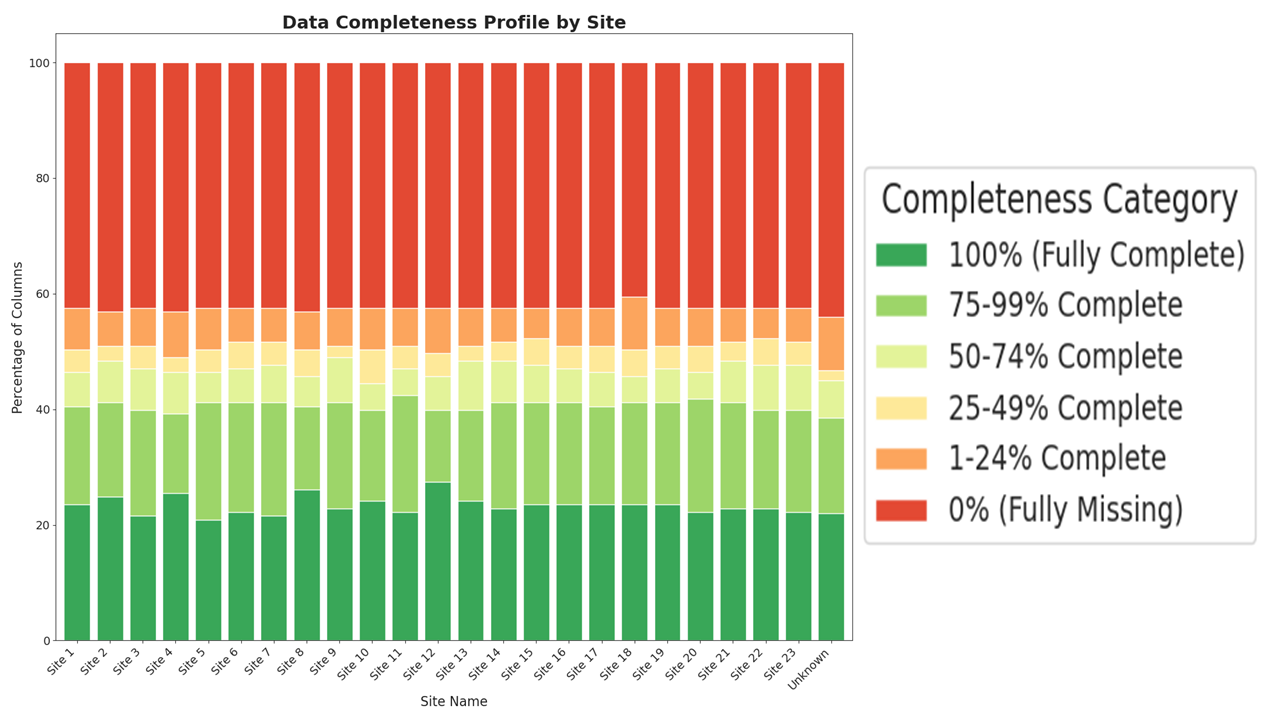}
    \caption{We calculated the completeness for each hospital site, and observed a slight difference in completeness.}
    \label{fig:rq3}
\end{figure}

We implemented several dimensions of the METRIC framework evaluation. We present the specific approaches and results we obtained for the different RQs:
    
For \textbf{\textit{RQ2 (Informative Missingness)}}, we assessed data completeness across different hospital sites, calculated for each column as the proportion of non-null, non-zero entries. We then compared these metrics across sites and data types. The divergence in data completeness between sources, illustrated in\textbf{Fig. \ref{fig:rq3}}, suggests that the data are not missing completely at random and that missingness is dependent on the data source.

\renewcommand{\arraystretch}{1.1}
    \begin{table}[htbp!]
    \centering
    \resizebox{\columnwidth}{!}{%
    \begin{tabular}{llll}
    \multicolumn{1}{l|}{}                              & \multicolumn{1}{l|}{Both ICD9 and ICD10} & \multicolumn{1}{l|}{ICD9 Only} & ICD10 Only \\ \cline{2-4} 
    \multicolumn{1}{l|}{\# unique concept mapped from} & \multicolumn{1}{l|}{7,125}               & \multicolumn{1}{l|}{7,754}     & 205,329    \\
    &&&           
    \end{tabular}%
    }
    \caption{Overlapping between ICD9-ICD10 and OMOP CDM codes}
    
    \label{tab:rq5}
    \end{table}
    
For \textbf{\textit{RQ3 (Timeliness)}}, we compared the overlap between ICD-9 and ICD-10 procedure codes. We found that only half of the ICD-9 codes shared a common mapping with an ICD-10 code (\textbf{Table \ref{tab:rq5}}). This limited overlap suggests that AI models are at risk of performance degradation when encountering data with different distributions of these coding systems. While ICD-10 codes were more prevalent, likely due to their greater comprehensiveness \cite{topaz2013icd}, the poor mapping indicates a broader potential risk that may affect other clinical vocabularies and warrants further investigation.

\begin{figure}[htbp!]
    \centering    
    \includegraphics[width=0.73\linewidth]{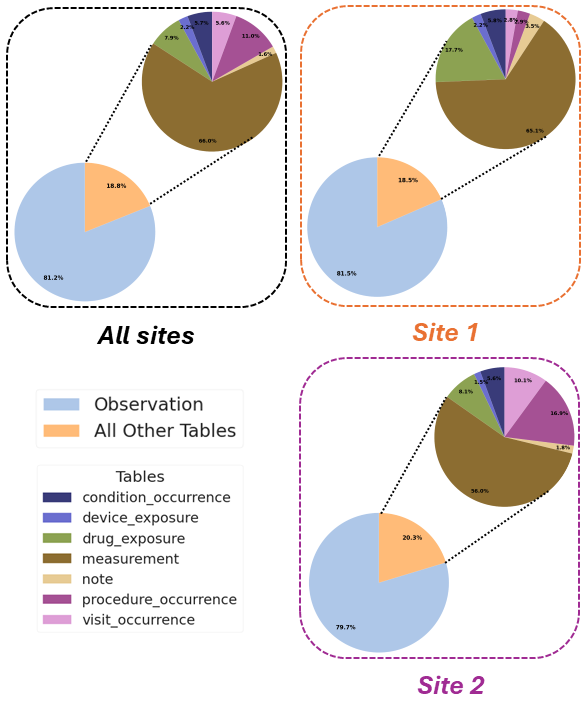}
    \caption{We represented the distribution of the different data for different data sources (hospital sites). We can observe that the distribution differs from one site to another.}
    \label{fig:rq6}
\end{figure}

For \textbf{\textit{RQ4 (Distribution Consistency)}}, we analyzed the distribution of different data types (e.g., procedures, conditions) across hospital sites. \textbf{Fig. \ref{fig:rq6}} shows that the data distributions varied between sites, likely reflecting different clinical specializations. However, consistent cross-site patterns were observed: observations (e.g., vital signs) were uniformly the most prevalent data type, while clinical notes were the least. We propose several hypotheses to explain this: (1) not all notes are entered into the EHR or may be omitted during the ETL process; (2) clinicians may input multiple billing codes for procedures to ensure accuracy, whereas notes may not be duplicated in the same way; and (3) a single note often summarizes multiple observations, procedures, or conditions, reducing overall note volume. We further hypothesize that clinical notes may implicitly contain content also represented as structured data in other categories.

\subsection{Case Study Specificity}

We identified key challenges and implementation barriers inherent to our use case. 
\subsubsection{Data Retrieval}
We retrieved patient data using CFM-specific ICD-9 and ICD-10 codes. However, we do not expect that a systematic OMOP CDM-based data quality assessment would have benefited data retrieval, as clinicians are familiar with ICD-9 and ICD-10 codes, but not necessarily with the OMOP CDM concept code. To nuance this, the OMOP CDM provides a centralized repository containing data from all hospital sites, which facilitates the retrieval of all CFM patients across all SC systems.
Moreover, the implementation of an automated cohort discovery tool (such as OHDSI Atlas) might greatly benefit data retrieval \cite{Atlas}. 

\subsubsection{Data Fidelity and Usefulness}
We observed that data fidelity and usefulness were more critical than their compliance with the OMOP CDM. For example, we identified limitations for systematic approaches regarding the assessment of data fidelity and usefulness for AI models. We think that a good evaluation of data usefulness would be represented by the upper bound of any AI model performance, meaning that it is case-study dependent.

\subsubsection{CFM Case-Study and FHIR Implementation}
\begin{figure}[htbp!]
    \centering    
    \includegraphics[width=0.9\linewidth]{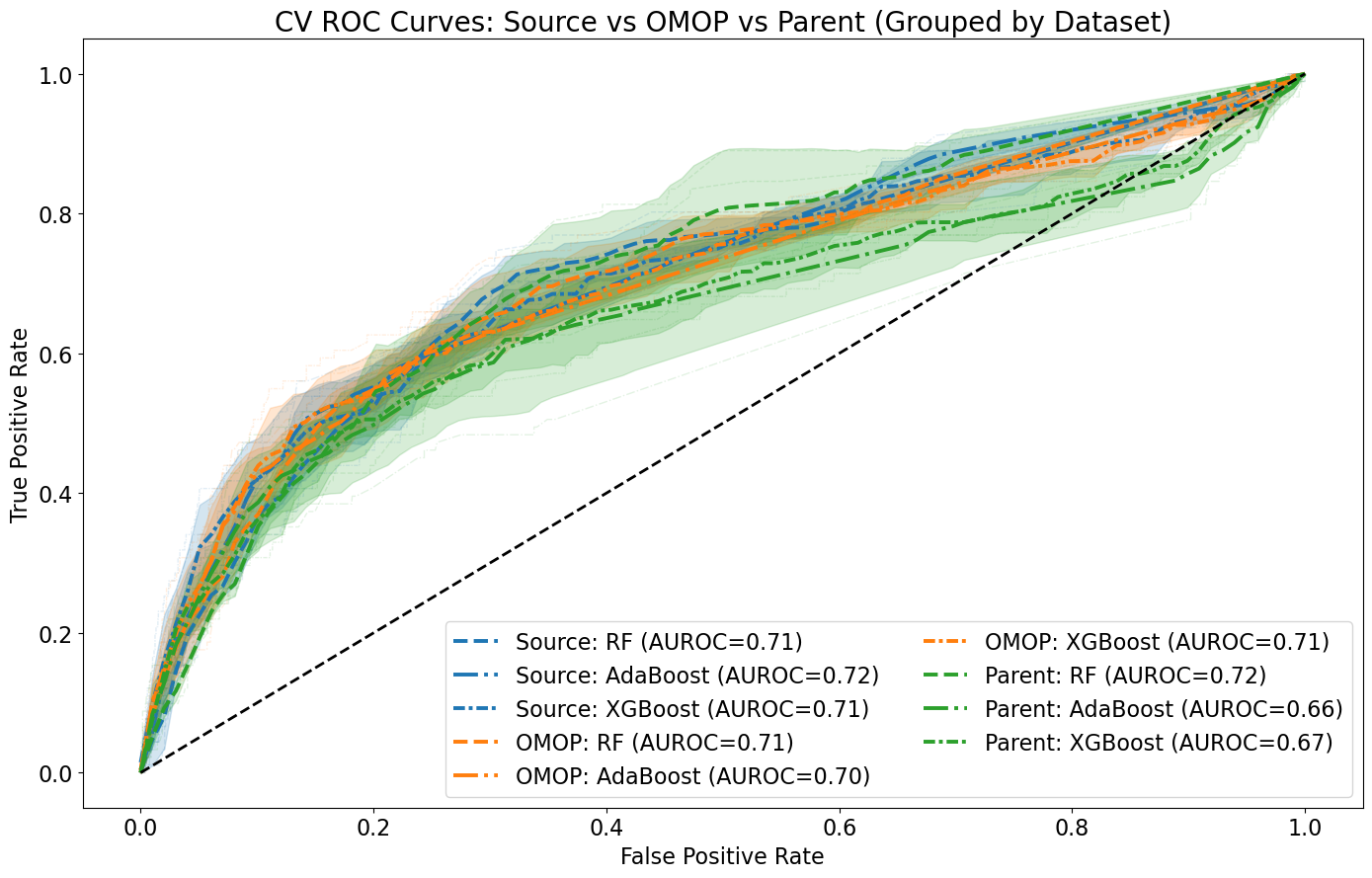}
    \caption{We show the AUROC for different models when using source codes (blue) vs. using harmonized OMOP CDM concept code (orange) vs. using supersets of OMOP CDM concept codes (green).}
    \label{fig:rq7}
\end{figure}

The CFM study enabled us to assess the impact of data harmonization on AI model performance. We used the Area Under the Receiver Operating Characteristic (AUROC) curve to evaluate our model performance, with the label being the patient's presence or absence of psychiatric diagnosis. We represent the results for 3 AI models: RandomForest (RF), eXtreme Gradient Boosting (XGBoost), and Adaptive Boosting (AdaBoost) \cite{breiman2001random, chen2016xgboost, freund1999short}., using 5-fold Cross-Validation (CV).
We observed two phenomenon based on \textbf{Fig. \ref{fig:rq7}}: (1) data harmonization do not significantly impact model performance with a mean AUROC of 71.3\% using source medical codes, and 70.0\% using OMOP CDM codes, validating \textbf{RQ5}, and (2) reducing the number of OMOP CDM concept code features decreased our AI model performance, suggesting that although making the taks "easier" (by reducing the number of features), we also loose granularity in the data. Further investigation is required as we did not validate \textbf{RQ6}, notably since we applied a brute-force approach in our AI model implementation. These results, however, are encouraging, as they provide additional evidence that data harmonization won't negatively impact AI model performance, while increasing interoperability and facilitating collaboration. We still believe that better approaches could leverage the OMOP CDM relationships to increase AI performance.

The implementation of FHIR within the SC MS Fabric environment presents significant challenges. SC MS Fabric is closed to external API connections, mitigating the implementation of FHIR servers capable of supporting FHIR Resource exchange. Moreover, FHIR does not rely on OMOP CDM concept code ID, but rather on the source codes (e.g., ICD-9, ICD-10), meaning that mapping from the OMOP CDM and FHIR is not straightforward, and might depend on the data context (e.g., specific ICD-9 codes map to different OMOP CDM concept code IDs, or vice-versa). This complexity limits the ability to implement FHIR systematically. However, we believe that FHIR remains the best option for implementing AI in case studies. Despite its limited scope, this work serves as a foundation by creating specific FHIR resources that support future research and development of FHIR infrastructure compatible with MS Fabric within SC's secure environment. 

\section{Discussions}
\label{Discussions}

The focus on AI implementation science in healthcare has increased in the last few years, with a positive impact on AI adoption by healthcare stakeholders (e.g., patients, providers, insurers) \cite{longhurst2024call}. Most importantly, this study demonstrates that implementing AI in a multisite healthcare system is not a trivial task and that systematic approaches are insufficient for the adoption of AI in healthcare. We believe that data distribution drift is one of the most critical risks that can compromise AI implementation, as it can significantly reduce AI performance (e.g., sudden population migration to one of the SC hospital sites) \cite{sahiner2023data}. Therefore, it is crucial to have an effective monitoring mechanism.
AI implementation is a complex task, often involving all stakeholders in a highly iterative development and implementation process. Based on our effort, we identified some critical variables that impact AI implementation.
\subsubsection{\textbf{Environmental}} The implementation depends on the resources available (e.g., cloud servers, computation capabilities), the existing infrastructure and workflows, which might impact usability of specific tools/ code (e.g., OHDSI DQD and Java within MS Fabric), and the type of interaction the end-user (e.g., clinicians) will have with the system. 
\subsubsection{\textbf{Data Access}} Data Access plays a significant role in AI Implementation. It depends on the local regulations in place, the existing data infrastructure, as well as standardization and harmonization. For example, access to raw data or access to only a subset will drive the scope and adoption of AI (e.g., restricted access to data will limit AI implementation).
\subsubsection{\textbf{Expertise Access}} As mentioned above, AI implementation is a collaborative enterprise, with many stakeholders with different expertise. However, the need from the primary end-users, such as patients and medical providers, will drive the need for implementation, while engineers and data scientists lead the technical implementation.

To contextualize our findings within medical data quality research, we benchmarked our experience against the large-scale European Health Data and Evidence Network (EHDEN) consortium \cite{blacketer2021using}. This comparison yields two insights for our study. First, our experience mirrors theirs in identifying data model conformance with the highest number of data quality issues, confirming that this is a common challenge in OMOP CDM implementation. Second, we observed a data quality gap, with the overall data quality scores improvement across the mature EHDEN network being higher than in our system implementation. We argue this gap is not a limitation but rather empirical evidence for the key implementation variables discussed previously, namely: \textbf{Environmental}, \textbf{Data Access}, and \textbf{Expertise}. Several key factors can explain the discrepancy. While there are apparent differences in the maturity of the data infrastructure, inherent variations in patient cohorts, and methodological differences can explain some of the differences, as the EHDEN analysis utilized a different and smaller subset of DQD tests (with an average of 881 tests per organization in the EHDEN study vs. 1798 tests performed in ours). This highlights that while general patterns are shared, a direct comparison of quality scores across institutions can be misleading without accounting for local context.

Ultimately, an ideal AI Implementation Science framework should encompass both systematic and case study needs, with a hybrid approach to AI implementation. Moreover, as demonstrated by the METRIC framework and TAI, AI Implementation Science researchers should draw inspiration from other fields with similar problems (e.g., finance for anomaly and data drift detection, or Extended Reality (XR) implementation frameworks in healthcare). In addition to TAI, AI Implementation Science should incorporate approaches and techniques from Safe AI, Actionable AI, or Responsible AI (STAR-AI).

\subsubsection*{Future Work}
There are several limitations to this study, notably the fact that we did not have access to the raw EHR data (pre-ETL process). This limits our ability to evaluate data fidelity, although we aim to apply AI-based techniques for improved data anomaly detection. We also plan to utilize clinical notes, combined with NLP techniques and AI models such as Large Language Models (LLMs), to obtain a secondary data source alongside the ETL process, thereby increasing confidence and trust in the data by reducing its uncertainty. As a next step, we will conduct a more formal comparison with external healthcare system, and perform a dedicated usability study of our tools, using the System Usability Scale (SUS) and open-ended surveys to gather feedback from end-users, namely SC's data engineers and physicians \cite{brooke1996sus}. Lastly, we plan to explore AI Implementation Science from the perspective of AI models.
\section{Conclusion}
\label{Conclusions}
This study highlights the methods and importance of integrating AI implementation science principles into the deployment of TAI within SC, a large, multi-site healthcare system. As AI technologies continue to advance, their capabilities will not be judged solely on accuracy, but also on their real-world impact. The integration of AI systems into clinical workflows, their acceptance by end-users, and their governance in accordance with ethical and regulatory standards will be key factors in determining the real-world clinical impact that AI implementation science addresses. Accordingly, our findings suggest that technical performance alone is not sufficient to ensure clinical utility; instead, attention must also be paid to factors such as data interoperability, clinician engagement, workflow alignment, and transparency of model outputs.

\section*{Internal Review Board Note}
For the systematic approach, the work was undertaken as a Quality Improvement Initiative at Shriners Hospitals for Children and, as such, was not formally supervised by an Institutional Review Board (IRB).

The CFM case study involving human subjects was conducted in accordance with the ethical standards outlined in the Belmont Report and received approval from the Georgia Institute of Technology, IRB approval number H21297.

\small
\bibliographystyle{unsrt}
\bibliography{Bibliography}

@misc{FHIR, key = {FHIR}, author={FHIR}, url={https://fhir.org/}, journal={HL7 FHIR Foundation Enabling health interoperability through FHIR}}

@misc{FABRIC, key = {Microsoft Fabric}, author={Microsoft}, url={https://www.microsoft.com/en-us/microsoft-fabric}, journal={Data Analytics Platform | Microsoft Fabric}}

@misc{Athena,
  author       = {{OHDSI Collaborative}},
  title        = {{OHDSI Athena}: Standardized Vocabularies},
  howpublished = {\url{https://athena.ohdsi.org/}},
  note         = {Accessed 2025}
}

@misc{Atlas,
  author       = {{OHDSI Collaborative}},
  title        = {{OHDSI Atlas}: Web-based Cohort and Analysis Tool},
  howpublished = {\url{https://atlas-demo.ohdsi.org/}},
  note         = {Accessed 2025}
}

@article{blacketer2021using,
  title={Using the data quality dashboard to improve the EHDEN network},
  author={Blacketer, Clair and Voss, Erica A and DeFalco, Frank and Hughes, Nigel and Schuemie, Martijn J and Moinat, Maxim and Rijnbeek, Peter R},
  journal={Applied Sciences},
  volume={11},
  number={24},
  pages={11920},
  year={2021},
  publisher={MDPI}
}

@article{hripcsak2015observational,
  title={Observational Health Data Sciences and Informatics (OHDSI): opportunities for observational researchers},
  author={Hripcsak, George and Duke, Jon D and Shah, Nigam H and Reich, Christian G and Huser, Vojtech and Schuemie, Martijn J and Suchard, Marc A and Park, Rae Woong and Wong, Ian Chi Kei and Rijnbeek, Peter R and others},
  journal={Studies in health technology and informatics},
  volume={216},
  pages={574},
  year={2015},
  publisher={NIH Public Access}
}

@article{acosta2022multimodal,
  title={Multimodal biomedical AI},
  author={Acosta, Juli{\'a}n N and Falcone, Guido J and Rajpurkar, Pranav and Topol, Eric J},
  journal={Nature medicine},
  volume={28},
  number={9},
  pages={1773--1784},
  year={2022},
  publisher={Nature Publishing Group US New York}
}

@incollection{bohr2020current,
  title={Current healthcare, big data, and machine learning},
  author={Bohr, Adam and Memarzadeh, Kaveh},
  booktitle={Artificial intelligence in healthcare},
  pages={1--24},
  year={2020},
  publisher={Elsevier}
}

@incollection{gerke2020ethical,
  title={Ethical and legal challenges of artificial intelligence-driven healthcare},
  author={Gerke, Sara and Minssen, Timo and Cohen, Glenn},
  booktitle={Artificial intelligence in healthcare},
  pages={295--336},
  year={2020},
  publisher={Elsevier}
}

@article{kumar2021data,
  title={Data harmonization for heterogeneous datasets: a systematic literature review},
  author={Kumar, Ganesh and Basri, Shuib and Imam, Abdullahi Abubakar and Khowaja, Sunder Ali and Capretz, Luiz Fernando and Balogun, Abdullateef Oluwagbemiga},
  journal={Applied Sciences},
  volume={11},
  number={17},
  pages={8275},
  year={2021},
  publisher={MDPI}
}

@article{chen2021personalized,
  title={Personalized health care and public health in the digital age},
  author={Ch{\'e}n, Oliver Y and Roberts, Bryn},
  journal={Frontiers in digital health},
  volume={3},
  pages={595704},
  year={2021},
  publisher={Frontiers Media SA}
}

@article{antunes2022federated,
  title={Federated learning for healthcare: Systematic review and architecture proposal},
  author={Antunes, Rodolfo Stoffel and Andr{\'e} da Costa, Cristiano and K{\"u}derle, Arne and Yari, Imrana Abdullahi and Eskofier, Bj{\"o}rn},
  journal={ACM Transactions on Intelligent Systems and Technology (TIST)},
  volume={13},
  number={4},
  pages={1--23},
  year={2022},
  publisher={ACM New York, NY}
}

@article{he2024foundation,
  title={Foundation model for advancing healthcare: challenges, opportunities and future directions},
  author={He, Yuting and Huang, Fuxiang and Jiang, Xinrui and Nie, Yuxiang and Wang, Minghao and Wang, Jiguang and Chen, Hao},
  journal={IEEE Reviews in Biomedical Engineering},
  year={2024},
  publisher={IEEE}
}

@article{alsaad2024multimodal,
  title={Multimodal large language models in health care: applications, challenges, and future outlook},
  author={AlSaad, Rawan and Abd-Alrazaq, Alaa and Boughorbel, Sabri and Ahmed, Arfan and Renault, Max-Antoine and Damseh, Rafat and Sheikh, Javaid},
  journal={Journal of medical Internet research},
  volume={26},
  pages={e59505},
  year={2024},
  publisher={JMIR Publications Toronto, Canada}
}

@article{you2025clinical,
  title={Clinical trials informed framework for real world clinical implementation and deployment of artificial intelligence applications},
  author={You, Jacqueline G and Hernandez-Boussard, Tina and Pfeffer, Michael A and Landman, Adam and Mishuris, Rebecca G},
  journal={NPJ Digital Medicine},
  volume={8},
  number={1},
  pages={107},
  year={2025},
  publisher={Nature Publishing Group UK London}
}

@inproceedings{el2025bridging,
  title={Bridging the Gap: From AI Success in Clinical Trials to Real-World Healthcare Implementation—A Narrative Review},
  author={El Arab, Rabie Adel and Abu-Mahfouz, Mohammad S and Abuadas, Fuad H and Alzghoul, Husam and Almari, Mohammed and Ghannam, Ahmad and Seweid, Mohamed Mahmoud},
  booktitle={Healthcare},
  volume={13},
  number={7},
  pages={701},
  year={2025},
  organization={MDPI}
}

@article{schwabe2024metric,
  title={The METRIC-framework for assessing data quality for trustworthy AI in medicine: a systematic review},
  author={Schwabe, Daniel and Becker, Katinka and Seyferth, Martin and Kla{\ss}, Andreas and Schaeffter, Tobias},
  journal={NPJ Digital Medicine},
  volume={7},
  number={1},
  pages={203},
  year={2024},
  publisher={Nature Publishing Group UK London}
}

@article{kahn2016harmonized,
  title={A harmonized data quality assessment terminology and framework for the secondary use of electronic health record data},
  author={Kahn, Michael G and Callahan, Tiffany J and Barnard, Juliana and Bauck, Alan E and Brown, Jeff and Davidson, Bruce N and Estiri, Hossein and Goerg, Carsten and Holve, Erin and Johnson, Steven G and others},
  journal={Egems},
  volume={4},
  number={1},
  pages={1244},
  year={2016}
}

@article{koscielniak2022shonet,
  title={The SHOnet learning health system: infrastructure for continuous learning in pediatric rehabilitation},
  author={Koscielniak, Nikolas and Jenkins, Diane and Hassani, Sahar and Buckon, Cathleen and Tucker, Joshua S and Sienko, Susan and Tucker, Carole A},
  journal={Learning Health Systems},
  volume={6},
  number={3},
  pages={e10305},
  year={2022},
  publisher={Wiley Online Library}
}

@article{reinecke2021usage,
  title={The usage of OHDSI OMOP--a scoping review},
  author={Reinecke, Ines and Zoch, Mich{\'e}le and Reich, Christian and Sedlmayr, Martin and Bathelt, Franziska},
  journal={German Medical Data Sciences 2021: Digital Medicine: Recognize--Understand--Heal},
  pages={95--103},
  year={2021},
  publisher={IOS Press}
}

@inproceedings{marteau2024improving,
  title={Improving A Large Healthcare System Research Data Warehouse Using OHDSI's Data Quality Dashboard},
  author={Marteau, Benoit L and Hornback, Andrew and Zhong, Yishan and Lowson, Christian and Woloff, Jason and Smith, Benjamin M and Hilton, Coleman and Wang, May D},
  booktitle={2024 IEEE EMBS International Conference on Biomedical and Health Informatics (BHI)},
  pages={1--8},
  year={2024},
  organization={IEEE}
}

@incollection{lima2019transforming,
  title={Transforming two decades of ePR data to OMOP CDM for clinical research},
  author={Lima, Daniel M and Rodrigues-Jr, Jose F and Traina, Agma JM and Pires, Fabio A and Gutierrez, Marco A},
  booktitle={MEDINFO 2019: Health and Wellbeing e-Networks for All},
  pages={233--237},
  year={2019},
  publisher={IOS Press}
}

@article{wang2021covid,
  title={COVID-19 SignSym: a fast adaptation of a general clinical NLP tool to identify and normalize COVID-19 signs and symptoms to OMOP common data model},
  author={Wang, Jingqi and Abu-el-Rub, Noor and Gray, Josh and Pham, Huy Anh and Zhou, Yujia and Manion, Frank J and Liu, Mei and Song, Xing and Xu, Hua and Rouhizadeh, Masoud and others},
  journal={Journal of the American Medical Informatics Association},
  volume={28},
  number={6},
  pages={1275--1283},
  year={2021},
  publisher={Oxford University Press}
}

@article{topaz2013icd,
  title={ICD-9 to ICD-10: evolution, revolution, and current debates in the United States},
  author={Topaz, Maxim and Shafran-Topaz, Leah and Bowles, Kathryn H},
  journal={Perspectives in Health Information Management/AHIMA, American Health Information Management Association},
  volume={10},
  number={Spring},
  pages={1d},
  year={2013}
}

@misc{longhurst2024call,
  title={A call for artificial intelligence implementation science centers to evaluate clinical effectiveness},
  author={Longhurst, Christopher A and Singh, Karandeep and Chopra, Aneesh and Atreja, Ashish and Brownstein, John S},
  journal={NEJM AI},
  volume={1},
  number={8},
  pages={AIp2400223},
  year={2024},
  publisher={Massachusetts Medical Society}
}

@article{sahiner2023data,
  title={Data drift in medical machine learning: implications and potential remedies},
  author={Sahiner, Berkman and Chen, Weijie and Samala, Ravi K and Petrick, Nicholas},
  journal={The British Journal of Radiology},
  volume={96},
  number={1150},
  pages={20220878},
  year={2023},
  publisher={Oxford University Press}
}

@article{wray2023bridging,
  title={Bridging the Know-Do Gap in Hospital Care Transitions},
  author={Wray, Charlie M and Jones, Christine D},
  journal={JAMA Internal Medicine},
  volume={183},
  number={5},
  pages={424--425},
  year={2023},
  publisher={American Medical Association}
}

@article{breiman2001random,
  title={Random forests},
  author={Breiman, Leo},
  journal={Machine learning},
  volume={45},
  number={1},
  pages={5--32},
  year={2001},
  publisher={Springer}
}

@inproceedings{chen2016xgboost,
  title={Xgboost: A scalable tree boosting system},
  author={Chen, Tianqi and Guestrin, Carlos},
  booktitle={Proceedings of the 22nd acm sigkdd international conference on knowledge discovery and data mining},
  pages={785--794},
  year={2016}
}

@article{freund1999short,
  title={A short introduction to boosting},
  author={Freund, Yoav and Schapire, Robert and Abe, Naoki},
  journal={Journal-Japanese Society For Artificial Intelligence},
  volume={14},
  number={771-780},
  pages={1612},
  year={1999},
  publisher={JAPANESE SOC ARTIFICIAL INTELL}
}

@article{brooke1996sus,
  title={SUS-A quick and dirty usability scale},
  author={Brooke, John and others},
  journal={Usability evaluation in industry},
  volume={189},
  number={194},
  pages={4--7},
  year={1996},
  publisher={London, England}
}

\end{document}